\documentstyle[12pt]{article}
\def\ifm{\ifmmode}

\newcommand{\notp}{\ \hbox{{$p$}\kern-.43em\hbox{/}}}
\newcommand{\notE}{\ \hbox{{$E$}\kern-.43em\hbox{/}}}
\newcommand{\beq}{\begin{eqnarray}}
\newcommand{\eeq}{\end{eqnarray}}
\newcommand{\beqs}{\begin{eqnarray*}}
\newcommand{\eeqs}{\end{eqnarray*}}

\def\PL #1 #2 #3 {{\it Phys. Lett.} {\bf#1} (#3) #2}
\def\NP #1 #2 #3 {{\it Nucl. Phys.} {\bf#1} (#3) #2}
\def\ZP #1 #2 #3 {{\it Z. Phys.} {\bf#1} (#3) #2}
\def\PRL #1 #2 #3 {{\it Phys. Rev. Lett.} {\bf #1} (#3) #2}
\def\PR #1 #2 #3 {{\it Phys. Rev.} {\bf#1} (#3) #2}
\def\MPL #1 #2 #3 {{\it Mod. Phys. Lett.} {\bf#1} (#3) #2}
\def\RMP #1 #2 #3 {{\it Rev.~Mod. Phys.} {\bf#1} (#3) #2}
\begin{document}
\thispagestyle{empty}
\baselineskip=14pt
\begin{flushright}
FSU--HEP--930816\\
MAD/TH/93-6\\
MAD/PH/788\\
hep--ph/9308370\\
August 1993
\end{flushright}
\vglue 1.cm
\begin{center}
{\large\bf The Charm Content of $W+1$~jet Events as a Probe of\\[2.mm]
the Strange Quark Distribution Function}
\vglue 1.5cm
U.~Baur$^a$, F.~Halzen$^b$, S.~Keller$^a$, M.~L.~Mangano$^c$,
K.~Riesselmann$^b$\\
\vglue 1.5cm
{$^a$\it Physics Department, Florida State University,
Tallahassee, Florida 32306, USA\\}
\vglue 3.mm
{$^b$\it Physics Department, University of Wisconsin,
Madison, Wisconsin 53706, USA\\}
\vglue 3.mm
{$^c$\it INFN, Scuola Normale Superiore and Dipartimento di Fisica,
Pisa, Italy\\}
\vglue 3.8cm
{\bf ABSTRACT}
\end{center}
\vglue 0.2cm
{\rightskip 3pc
\leftskip 3pc
\noindent
We investigate the prospects for measuring the strange quark distribution
function of the proton in associated $W$ plus charm quark
production at the Tevatron. The $W+c$
quark signal produced by strange quark -- gluon fusion, $sg\rightarrow
W^-c$ and $\bar sg\rightarrow W^+\bar c$,
is approximately 5\% of the inclusive $W+1$ jet cross section for
jets with a transverse momentum $p_T(j)>10$~GeV. We study the
sensitivity of the $W$ plus charm quark cross section to the
parametrization
of the strange quark distribution function, and evaluate the various
background processes. Strategies to identify charm quarks in CDF and
D\O \ are discussed. For a charm tagging efficiency of about 10\%
and an integrated luminosity of 30~pb$^{-1}$ or more, it should be
possible to constrain the strange quark
distribution function from $W+c$ production at the Tevatron.
}
\newpage

\section{Introduction}

One of the main goals of deep inelastic scattering experiments is
to obtain reliable measurements of the parton distribution functions
of the proton.
Recently, the CTEQ~\cite{CTE93} and CCFR~\cite{CCF92} collaborations have
determined the ratio of momentum fractions of strange
quarks versus $\bar u$ plus $\bar d$ quarks,
\begin{eqnarray}
\noalign{\vskip 5pt}
\kappa ={2S\over\overline U+ \overline D}~,
\end{eqnarray}
where
\begin{eqnarray}
\noalign{\vskip 5pt}
Q=\int_0^1xq(x)dx
\end{eqnarray}
with $Q=\overline U,\, \overline D,\, S$ and $q(x)$ being the parton
distribution function of the $\bar u, \, \bar d$ and $s$-quark,
respectively.
The CCFR collaboration obtained $\kappa\approx 0.4$ in its
analysis of the process $\nu N\rightarrow\mu^+\mu^-X$,
whereas the CTEQ collaboration in its global fit found $\kappa\approx 1$.
The statistical significance of the
difference between the two results is approximately $3-4$ standard
deviations. The CTEQ result can be traced back to the
appropriately weighted difference of
$F_2^{\nu}$, measured by CCFR, and $F_2^{\mu}$ as determined by NMC,
which is proportional to the $s$-quark distribution function (up to
${\cal O}(\alpha_s)$ corrections); see Ref.~\ref{CTE93} for more details.
The CCFR result of $\kappa\approx 0.4$, obtained from the
di-muon analysis, is best represented by
the MRSD0 parametrization~\cite{MRS93}. The value of $\kappa$ in this
set was fixed at a value of $0.5$ at
$Q^2=4$~GeV$^2$. As a representative of the CTEQ result,
the CTEQ1M set will be used in the following.
\vskip 14pt

The large difference in $\kappa$ results in substantially different
$s$-quark distribution functions for the two parametrizations. This is
demonstrated in Fig.~1, where the ratio of the strange quark
distribution functions of the two sets is shown for $Q^2= 5$~GeV$^2$ and
$Q^2=M_W^2$. At small $Q^2$ and $x < 0.1$, the two $s$-quark
distribution functions differ
approximately by a factor of two. At large $Q^2$, the ratio of the two
sets
is closer to unity. This is due to the fact that additional strange
quark pairs originating from gluon splitting contribute to the strange
quark distribution function.  Because both sets have very similar gluon
distribution functions, the additional strange quark contribution is
almost
identical for both sets, and hence the relative difference of the two
distribution functions diminishes.
\vskip 14pt

The discrepancy between the CTEQ and MRS parametrizations could be
resolved by a direct and independent measurement of the strange quark
distribution function. In this letter it is suggested that such a
measurement
could be carried out at the Tevatron by determining the charm content
of $W+1$~jet events. Our paper is organized as follows. In
Section~2, the signal for constraining the strange quark distribution function
and the leading background processes are studied.  In Section~3,
different experimental techniques to tag a charm quark inside a jet are
discussed and the minimum integrated luminosity necessary to
discriminate between the CTEQ and MRS strange quark distribution function
is estimated. Finally, in Section~4, our conclusions are presented
together
with some additional remarks. Some preliminary results of the work
described here were presented in Ref.~\cite{BHKMR}.

\section{Signal and Background}

Associated $W+$~charm production proceeds, at lowest order, through $sg$
and $\bar sg$ fusion, $sg\rightarrow W^-c$ and $\bar sg\rightarrow
W^+\bar c$. The alternative process where the $s$-quark in the reaction is
replaced by a $d$-quark, is suppressed by the quark mixing matrix
element $V_{cd}$. This suppression is somewhat compensated by the
larger $d$ quark
distribution function, such that the $dg\rightarrow Wc$ cross section is
about 10\% of the $sg\rightarrow Wc$ rate.  Since the final state is
identical for these two subprocesses, the sum of the $dg$ and $sg$
contributions will be considered as the ``signal''.
\vskip 14pt

The potentially largest background originates from the production of a
$c\bar c$ pair in the jet recoiling against the $W$.  When only the $c$
or the $\bar c$ is identified in
the jet, such a $W+c\bar c$ event looks like a signal event.
Similarly, a $b\bar b$ pair can be produced in the jet, and the $b$ or the
$\bar b$-quark misidentified as a charm quark.
\vskip 14pt

For our subsequent discussion it is useful to define the following
ratios of cross sections:
\begin{eqnarray}
{\cal R}_1 & = & \frac{\sigma(signal)}{\sigma(W+1~jet)}~,\\ [2.mm]
R_1(p_T) & = & \frac{d\sigma(signal)/dp_T(j)}{d\sigma(W+1~jet)/dp_T(j)
}~, \\[2.mm]
{\cal R}_2 & = & \frac{\sigma(signal)+\sigma(bgd)}{\sigma(W + 1~jet)}~,
\\[2.mm]
R_2(p_T) & = & \frac{d\sigma(signal)/dp_T(j)+d\sigma(bgd)
/dp_T(j)}{d\sigma(W+1~jet)/dp_T(j)}~,
\end{eqnarray}
and
\beq
{\cal R}_3 = {\cal R}_2-{\cal R}_1, \hskip 1.cm R_3(p_T) = R_2(p_T)-R_1(p_T),
\eeq
where ``bgd'' includes the background processes mentioned earlier,
assuming that all $b$ and $\bar b$-quarks in $W+b\bar b$
production are misidentified as charm quarks.
The notation ``W + 1~jet'' refers to the total inclusive W + 1~jet cross
section within cuts. ${\cal R}_1$ represents the ideal situation in which
all
background events have been completely eliminated.
${\cal R}_2$ describes the conservative case where none of the leading
background processes is reduced. Finally, ${\cal R}_3$ is useful to
explore the
sensitivity of the background to differences between the CTEQ and MRS
sets.  The background processes receive contributions from quark --
antiquark and quark gluon fusion, with all the quarks
contributing. They are thus sensitive to global differences in
the CTEQ and MRS
sets, rather than to differences in the strange quark distribution functions.
\vskip 14pt

In practice, the ratio of events with a tag for a charm quark inside the
jet
over the total number of events will be measured.
The result will fall in between ${\cal R}_1$ and ${\cal R}_2$,
because specific methods of tagging the charm quark inside of the jet will
suppress some parts of the background (see Section~3). Compared to the
absolute cross sections, the ratios ${\cal R}_i$, i=1,2,3, have a number
of advantages. Many experimental uncertainties, for example the
uncertainty in the integrated luminosity, are expected to cancel, at
least partially, in the ratios. Furthermore, the sensitivity to the
factorization scale $Q^2$ is reduced in the cross section ratios.
\vskip 14pt

To numerically simulate the signal and background processes we use
the Monte--Carlo program PYTHIA~\cite{PYT87} (version~5.6).
All processes are studied at the parton level, {\it i.e.} final state
showers are included but fragmentation is not. The $c$ and $b$-quark
masses are taken to be $m_c=1.35$~GeV and $m_b=5$~GeV. For
$W+c\bar c$ and $W+b\bar b$ production, the result of PYTHIA was
compared with that of the matrix element calculation of Ref.~\cite{Man92}.
The
results of both calculations are in general agreement, with PYTHIA
resulting
in somewhat larger cross sections for the background. In our simulations,
a ``jet'' is defined as follows. The direction of the sum of the momenta
of all the partons produced in the shower is taken as the center of a cone
of radius $\Delta R_{jet}=
\sqrt{\Delta\eta^2 +\Delta\phi^2} = 0.7$, where $\eta$ is the
pseudorapidity
and $\phi$ the azimuthal angle. All the partons inside that cone are
considered part of the jet.  Only events with a charm quark {\it inside}
the jet cone are counted. Background
events with two charm quarks inside  the jet cone are counted twice.
Only about one half of the background events has two charm
quarks inside the cone.
\vskip 14pt

The $W^\pm$ is assumed to decay into a $e^\pm\nu$ final state. To simulate
the acceptance of a real detector,
the following transverse momentum and pseudorapidity cuts on the
final state particles are imposed:
\begin{eqnarray}
\label{eq:cut}
p_{T}(e)  & \geq & 20~{\rm GeV}, \hskip 1.cm |\eta(e)| \leq 1,  \nonumber
\\
\notp_{T} & \geq & 20~{\rm GeV}, \hskip 1.cm \phantom{|\eta(e)| \leq 1.}
\\
p_{T}(j) & \geq & 10~{\rm GeV}, \hskip 1.cm |\eta(j)| \leq 1. \nonumber
\end{eqnarray}
The ratios ${\cal R}_i$ do not depend sensitively on the cuts imposed on
the
$W$ decay products, and the jet pseudorapidity cut
chosen. They are, however, sensitive to the $p_T$ cut on the jet.
\vskip 14pt

Figure~2 shows the differential cross section, $d\sigma/dx_s$,
of the process $sg\rightarrow Wc$ for $p\bar p$
collisions at $\sqrt{s}=1.8$~TeV as a function of the momentum fraction of
the
strange quark, $x_s$, using the CTEQ1M parametrization and the cuts
listed in Eq.~\ref{eq:cut}.
$W$ plus charm quark production at the
Tevatron thus is sensitive to the strange quark distribution mostly in
the $x_s$ region between 0.04 and 0.1, in which the CTEQ and MRS
parametrizations are indeed substantially different (see Fig.~1).
\vskip 14pt

The cross sections for the signal, the various background processes, and
inclusive $W+1$~jet production at the Tevatron are given in
Table~\ref{tab:xsect}.  Approximately 75\% (20\%) of the
background originates from a $c\bar c$ ($b \bar b$) pair produced in a jet
initiated by a gluon, if we assume that all $b$ and $\bar b$-quarks are
misidentified as charm quarks.
The remaining 5\% is due to the production of a $c \bar c$ pair in a
quark--initiated jet. The combined background cross section is about equal
to the signal rate. Numerical values for ${\cal R}_1$, ${\cal R}_2$ and
${\cal R}_3$ are presented in Table~\ref{tab:rat} for the two sets of
parton distribution functions.
The signal accounts approximately for 4--5\% of all $W+1$~jet events
(${\cal R}_1$), and the background for about 4\% (${\cal R}_3$).
As can be seen from Table~\ref{tab:xsect}, the two sets of parton
distribution
functions yield the same values for the inclusive $W+1$~jet cross section
and
the three background processes to within 2\%.  Correspondingly, the
variation of ${\cal R}_3$ with the two sets is small.
The signal rate, on the other hand, is quite sensitive to which set is
chosen, as expected.  The variation of the ratios ${\cal R}_1$ and ${\cal
R}_2$
directly reflects the difference in the strange quark distribution.
\vskip 14pt

Not surprisingly, the cross section ratios are relatively insensitive to
changes in the factorization scale $Q^2$.
Varying $Q^2$ between 1/4 and 4 times the default average $Q^2$ of PYTHIA,
the ratios change only by $\Delta {\cal R}_i/{\cal R}_i\approx4$\%,
although
individual cross sections vary by up to 20\% . The stability of ${\cal
R}_1$
and ${\cal R}_2$ with respect to variations in $Q^2$ indicates that the
sensitivity
of the ratios to the strange quark distribution function is unlikely to be
overwhelmed by uncertainties originating from higher order QCD
corrections.
The signal cross section
is insensitive to variations in the $c$-quark mass, due to the high $p_T(j)$
cut used.  The background is slightly sensitive to changes in the $c$
and $b$-quark
masses, resulting in a $\sim 2\%$ variation in ${\cal R}_2$ when $m_c$
($m_b$) is changed from 1.5~GeV (5~GeV) to 1.35~GeV (4.5~GeV).
\vskip 14pt

We have also investigated the sensitivity of our results to the jet cone
size $\Delta R_{jet}$. Reducing the cone size from 0.7 to 0.4, the
signal, background and inclusive $W+1$~jet cross sections are reduced
by approximately the same amount for both sets of parton distribution
functions (10\% for the signal, 40\% for the
background, and 20\% for the inclusive $W+1$~jet rate).
A change in $\Delta
R_{jet}$ therefore does not alter the sensitivity of ${\cal R}_1$ and
${\cal
R}_2$ to the strange quark distribution function in first approximation.
This result can be easily understood by the general insensitivity of the
background and inclusive $W+1$~jet cross sections to the choice of
parton distribution functions, and by noticing that the ratio of the two
strange quark
distribution functions varies only little in the $x_s$ range accessible
in $sg\rightarrow Wc$ (see Figs.~1 and~2). For a constant ratio of
$s$-quark distribution functions, the relative change of the cross
section ratios would be independent of the jet cone size.
\vskip 14pt

The differential cross section ratios $R_1(p_T)$, $R_2(p_T)$, and
$R_3(p_T)$ as a function of the jet
transverse momentum are displayed in Fig.~3. Due to the different $x$
behavior of parton distribution functions, the ratios slowly grow with
$p_T(j)$. The difference in $R_1(p_T)$ and $R_2(p_T)$ for the two
sets of parton
distribution functions is fairly uniform in the jet transverse momentum.
$R_3(p_T)$ is practically insensitive to parton distribution function effects
over the whole jet $p_T$ range studied. Figure~3 demonstrates that
the $W$ plus charm cross section is sensitive to the $s$-quark
distribution
function over the whole jet transverse momentum range.
Of course, the region of low $p_T(j)$ will give better statistics.
\vskip 14pt

We can now estimate the minimum charm tagging
efficiency, $\epsilon^{min}_c$, required to be statistically sensitive to
the
variation of the $Wc$ production cross section with the strange quark
distribution function. To quantify the difference between the two sets
of parametrizations in the ratios ${\cal R}_i$, the following variable is
used:
\beq
\noalign{\vskip 5pt}
\Delta_i=2~\frac{{\cal R}_i(CTEQ1M)-{\cal R}_i(MRSD0)}{{\cal
R}_i(CTEQ1M)+{\cal R}_i(MRSD0)}\, .
\eeq
Using the numbers listed in Table~\ref{tab:rat}, we find
$\Delta_1=25\%$,  $\Delta_2=14\%$, and $\Delta_3=2\%$.
Assuming both electron and muon decay channel of the $W^{\pm}$ boson,
an integrated luminosity of 10~pb$^{-1}$ yields about 1700
$W+1$~jet events for the cuts described in Eq. 8. This corresponds to
approximately 75 -- 90 $W$ plus charm quark signal events, and to about
the
same number of potential background events. To experimentally
differentiate between the two sets MRSD0 and CTEQ1M, the experimental
uncertainties in measuring ${\cal R}_1$ (${\cal R}_2$) must be less or
equal to $\Delta_1$ ($\Delta_2$). From the expected number of
signal events it is straightforward to estimate $\epsilon^{min}_c$.
Depending on how efficiently the various background processes can be
suppressed, an efficiency $\epsilon^{min}_c\approx 20 - 30$\%
for an integrated luminosity of 10~pb$^{-1}$ is needed.
Note that $\epsilon^{min}_c$ scales like $(\int\!{\cal L}dt)^{-1}$.
\vskip 14pt

\section{Charm Quark Tagging in Tevatron Experiments}

The two collider experiments, CDF and D\O, at the Tevatron explore
three different strategies to identify charm quarks:

\begin{enumerate}

\item Search for a displaced secondary vertex in the silicon vertex
detector (SVX). The efficiency to tag $b$-quarks with the SVX~\cite{SVX90}
of
CDF is about 10--20\%, depending on the $p_T$ range.
The tagging efficiency for the charm quarks is
expected to be smaller than that for bottom quarks as a result of the
smaller
mass and decay track multiplicity of the charmed hadrons.

\item Reconstruction of exclusive nonleptonic charmed baryon or
meson decays.
CDF, for example, uses the decay channel $D^0 \rightarrow K \pi$ to
identify semileptonic B meson decays~\cite{CDF93}.
Other exclusive channels will be added in the future, and an efficiency of
a
few percent should be reached.

\item Looking for inclusive semileptonic charm decays~\cite{LEP93}.
The average inclusive semileptonic charm decay branching ratio is
$B(c\rightarrow
e\nu,\mu\nu)\sim 10$\%. If one assumes a reconstruction efficiency for a
muon inside a jet of the order of 50\%~\cite{BLAZ}, a total charm
tagging efficiency
from semileptonic charm decays of the order of 5\% may well be possible.
\end{enumerate}

Combined, the three methods may yield an overall charm detection
efficiency of about 10\%. Based on this assumption, an integrated
luminosity
of ${\cal O}(30$~pb$^{-1})$ should provide the first statistically
significant information on the strange quark distribution of the proton.
\vskip 14pt

A more precise estimate of the minimum integrated luminosity required
depends on a better understanding of the charm quark detection
efficiency, and on more detailed background studies. In
principle, the three background processes considered here can be
reduced by:

\begin{itemize}
\item Charge reconstruction: for the signal, the $W$ and $c$
quark electric charges are correlated. For the $c\bar c$ background, the
charm quark has the wrong charge 50\% of the time. Therefore, if the
charges
of
the $W$ and of the charm quark can be determined, the $Wc\bar c$
background can be reduced by
a factor of two.  Furthermore, events with the wrong charge correlation
provide a measurement of the background, that could subsequently be
subtracted.

\item Cut on the charm transverse momentum: since more than
one charm quark is present in the background processes its average $p_T$
is smaller than in the signal. This is illustrated in Fig.~4 where
the $c$-quark transverse momentum distribution for the signal
(solid histogram) and the background (dashed histogram) is shown, using
the
CTEQ1M set of parton distribution functions. Due to the
$p_T(j)>10$~GeV cut, the $p_T$ distribution of the charm quark in the
signal sharply peaks at a value of about 10~GeV. On the other hand, the
transverse momentum distribution of charm quarks originating from the
background processes considered, peaks at $p_T\approx 5$~GeV.

\item Flavor identification: if the bottom quark is
identified, the $b\bar b$ background can be subtracted.
\end{itemize}

\section{Conclusions}

We have studied the prospects for constraining the
strange quark distribution function in $W+c$ production at the Tevatron.
The method we suggest is similar to the one described in Ref.~\cite{Fle89}
for measuring the charm quark distribution function in $\gamma$ plus
charm  production.
Our results indicate that, for the data sample accumulated in the
1992-93 run, the expected charm tagging efficiencies are a limiting
factor.
For an integrated luminosity of 30~pb$^{-1}$, a charm tagging efficiency
of
at least 10\% is needed.
However, with an integrated luminosity of about 100~pb$^{-1}$,
which is expected by the end of 1994, it should be possible to make a
serious attempt at
constraining the strange quark distribution function from $W+c$
production.
\vskip 14pt

In our analysis we have concentrated on the charm content of $W+1$~jet
events with $p_T(j) > 10$~GeV. Alternatively one could search for $W+c$
production in the inclusive $W$ sample, without requiring the presence
of a high transverse momentum jet. The advantage here would be a
significant increase of the number of signal events. However,
due to the smaller average transverse momentum of the charm quarks
in the inclusive $W$ sample, the charm quark detection efficiency is
expected to be reduced.
Furthermore, the ratio of signal to background will decrease.
\vskip 14pt

The largest uncertainties in our calculation arise from the overall
charm tagging efficiency at CDF and D\O, and from uncertainties in the
estimate of potential background processes.
Clearly, more experimental and theoretical work is needed.
\vskip 14pt

If the strange quark distribution function is measured precisely in
other experiments, $W$ plus charm quark production may eventually be used
to measure the quark mixing matrix element $V_{cs}$ at high $Q^2$ and
compare it with the value extracted from low energy
experiments~\cite{LIN}.
\vskip 14pt

\section{Acknowledgements}

We are grateful to G.~C.~Bla\-zey,
S.~Erre\-de, B.~Flaugher, B.~Fletcher, T.~Heuring, T.~LeCompte,  H.~Reno,
and
M.~L.~Stong for useful and stimulating discussions.  This research was
supported in part by the  Texas National
Research Laboratory Commission under grant no. RGFY9273, and by the
U.S. Department of Energy under
contract numbers DE-FG05-87ER40319, and DE-AC02-76ER00881.

\newpage

\begin{table}[h]
\centering
\caption
[Cross sections.]
{Cross sections for associated $W$ plus charm quark production with
$W\rightarrow e\nu$ at the Tevatron, using the MRSD0 and CTEQ1M
parametrizations of the parton distribution functions.
The cuts imposed are listed in Eq.~8.}
\vspace{6.mm}
\begin{tabular}{|c||c|c|c|c|c|}\hline
        & \multicolumn{5}{c|}{Cross section (pb)} \\ \cline{2-6}
	& Signal & \multicolumn{3}{c|}{Background} & Inclusive \\
\cline{2-6}
	& $ c\rightarrow c + X$ & $g\rightarrow c\bar c + X$
&$g\rightarrow
b\bar b + X$
			&$q\rightarrow qc\bar c + X $& $W$+ 1 jet \\
\hline\hline
MRSD0  & 3.68 & 2.76 & 0.75 & 0.15 & 87.0 \\ \hline
CTEQ1M & 4.58 & 2.80 & 0.77 & 0.15 & 85.6 \\ \hline
\end{tabular}
\label{tab:xsect}
\end{table}
\vglue 2.cm

\begin{table}[h]
\centering
\caption
[Ratios.]
{The ratios ${\cal R}_1$, ${\cal R}_2$ and ${\cal R}_3$ for the MRSD0 and
CTEQ1M parametrization. The cuts imposed are listed in Eq.~8.
}
\vspace{4.mm}
\begin{tabular}{|c||c|c|c|}\hline
&${\cal R}_1$& ${\cal R}_2$ & ${\cal R}_3$ \\ \hline\hline
MRSD0  &0.042    &0.084  & 0.042  \\ \hline
CTEQ1M &0.054    &0.097  & 0.043  \\ \hline
\end{tabular}
\label{tab:rat}
\end{table}

\newpage

\begin{figure}[h]
\vskip 3.3in
%\special{psfile=fig1.ps hoffset=0  voffset=-200 hscale=80 vscale=80 angle=0}
\caption{Ratio of the strange quark distribution functions as a
function of $x$ for the
CTEQ1M and MRSD0 sets and two different values of $Q^2$.}
\end{figure}

\begin{figure}[h]
\vskip 3.3in
%\special{psfile=fig2.ps hoffset=0 voffset=-200 hscale=80 vscale=80 angle=0}
\caption{The differential cross section $d\sigma/dx_s$ for the process
$sg\rightarrow Wc$ at the Tevatron, using the CTEQ1M set.
The cuts imposed are listed in Eq.~8.}
\end{figure}

\newpage
\begin{figure}[h]
\vspace{3.3in}
%\special{psfile=fig3.ps hoffset=0. voffset=-200 hscale=80 vscale=80 angle=0}
\caption{The differential cross section ratios $R_1(p_T)$,
$R_2(p_T)$, and $R_3(p_T)$ as a function of the
jet $p_T$ for the MRSD0 and CTEQ1M sets at the Tevatron.
The cuts imposed are listed in Eq.~8.}
\end{figure}
\begin{figure}[h]
\vspace{3.3in}
%\special{psfile=fig4.ps hoffset=0 voffset=-200 hscale=80 vscale=80 angle=0}
\caption{The transverse momentum distribution of the charm quark inside
the
jet for the signal, and the three background processes combined, at the
Tevatron, using the CTEQ1M set.  The cuts imposed are
listed in Eq.~8.}
\end{figure}

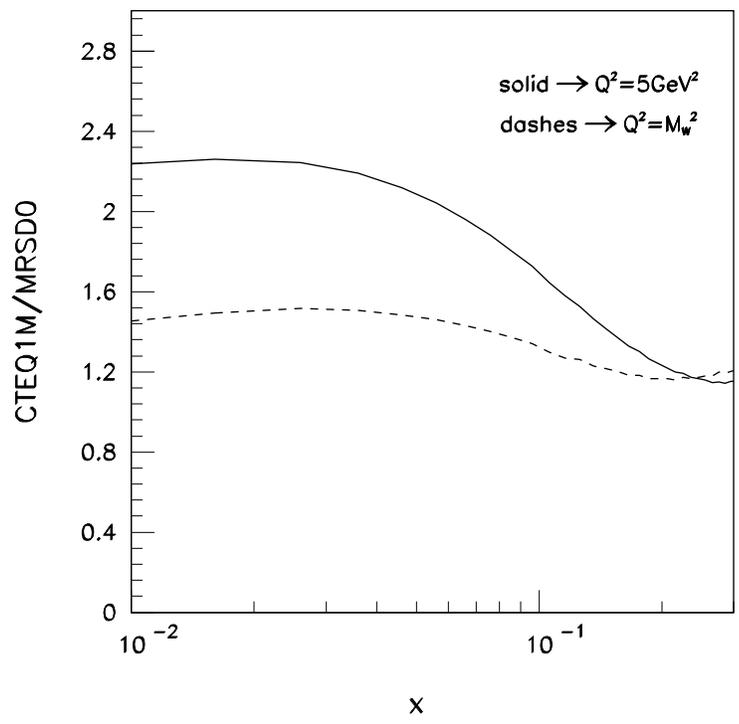
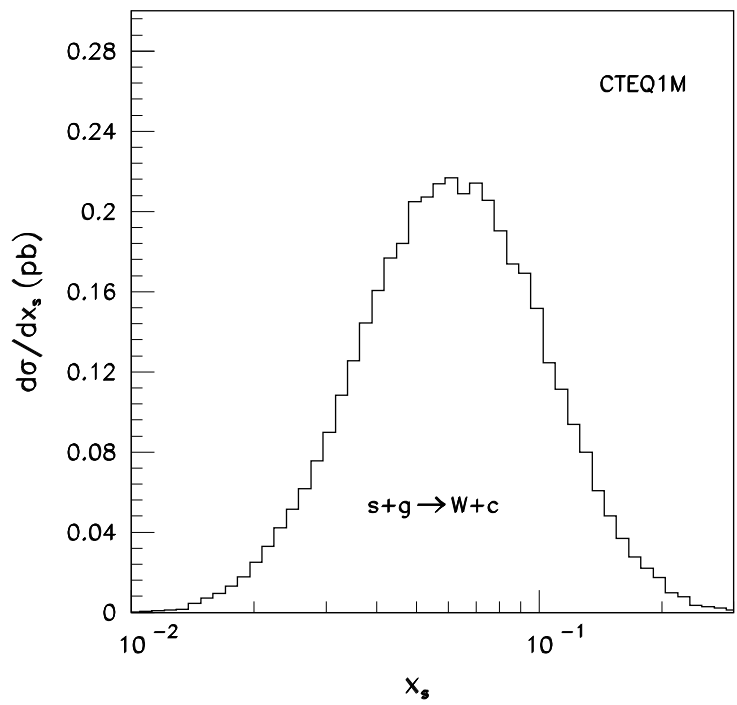
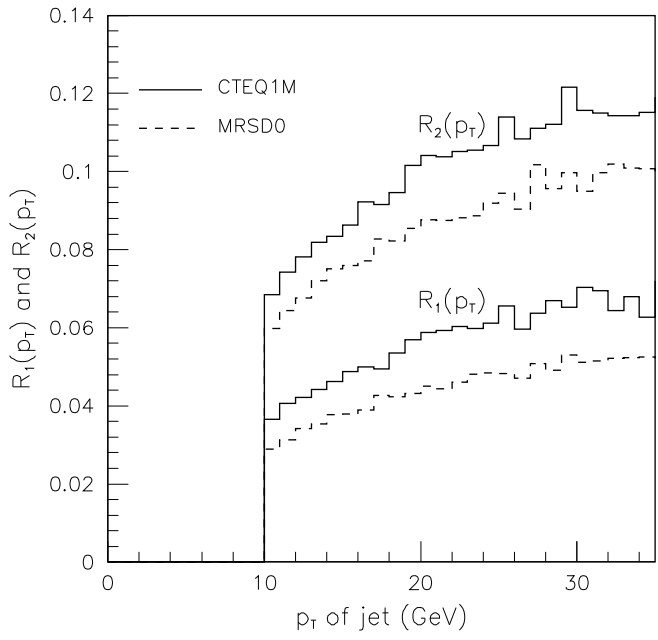
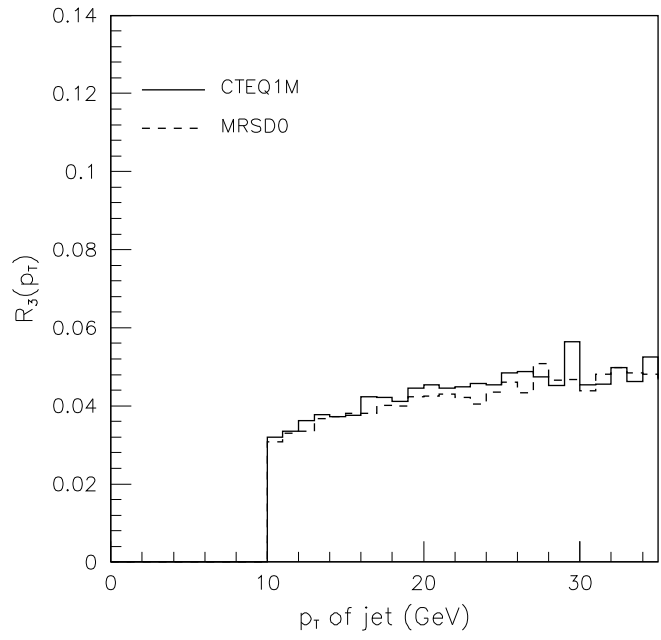
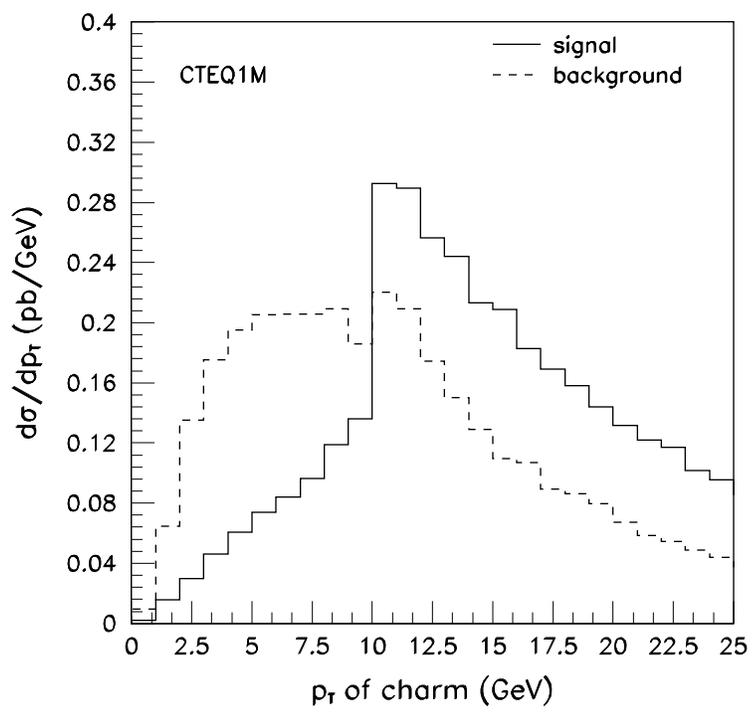
\end{document}